\documentclass{article}
\usepackage{spconf,amsmath,graphicx,hyperref}
\usepackage{booktabs}
\usepackage[dvipsnames]{xcolor}
\usepackage{multirow}
\usepackage{caption}
\usepackage{subcaption}

\title{A Semantic Timbre Dataset for the Electric Guitar}
\name{Joseph M. Cameron, Alan F. Blackwell\thanks{The authors thank and acknowledge the Department of Computer Science, University of Cambridge \& the UKRI EPSRC for funding this research.}}
\address{Department of Computer Science \& Technology, University of Cambridge\\Cambridge, United Kingdom}
\begin{document}
%\ninept
%
\maketitle
\begin{abstract}
Understanding and manipulating timbre is central to audio synthesis, yet this remains under-explored in machine learning due to a lack of annotated datasets linking perceptual timbre dimensions to semantic descriptors. We present the \textbf{Semantic Timbre Dataset}, a curated collection of monophonic electric guitar sounds, each labeled with one of 19 semantic timbre descriptors and corresponding magnitudes. These descriptors were derived from a qualitative analysis of physical and virtual guitar effect units and applied systematically to clean guitar tones. The dataset bridges perceptual timbre and machine learning representations, supporting learning for timbre control and semantic audio generation. We validate the dataset by training a variational autoencoder (VAE) on its latent space and evaluating it using human perceptual judgments and descriptor classifiers. Results show that the VAE captures timbral structure and enables smooth interpolation across descriptors. We release the dataset, code, and evaluation protocols to support timbre-aware generative AI research.
\end{abstract}
\begin{keywords}
Acoustic Signal Processing, Generative AI, Dataset Creation, Timbre, Dataset Evaluation
\end{keywords}
%

%%%%%%%%%%%%%%%%%%%%%%%%
%----------------------%
%%%%%%%%%%%%%%%%%%%%%%%%

\section{Introduction}
\label{sec:intro}

Timbre, the perceptual quality that distinguishes sounds of identical pitch and loudness, is fundamental to audio synthesis and musical expression \cite{siedenburg_present_2019}. Yet despite its importance \cite{liu_emotional_2018,holmes_exploration_2012}, timbre remains difficult to model and control in generative machine learning systems, primarily due to a lack of annotated datasets that map perceptual qualities to semantic descriptors. Existing corpora such as NSynth \cite{engel_nsynth_2017} offer large-scale audio data but lack rich, high-resolution semantic annotations, severely limiting models’ capacity for intuitive control over timbre.

Semantic descriptors like `bright', `fuzzy', or `smooth' are widely used by musicians and producers \cite{siedenburg_semantics_2019} to shape audio through effects pedals and digital Virtual Studio Technology (VST) plugins. However, current generative models typically require low-level parameter adjustments that do not align with how users conceptualize sound \cite{stables_safe_2014}. Bridging this semantic gap requires datasets that explicitly link these perceptual terms to acoustic properties.

While recent work has explored semantic descriptors in audio modeling, most prior approaches lack systematically annotated datasets with fine-grained perceptual control. For instance, NSynth \cite{engel_nsynth_2017} provides large-scale audio data but includes only coarse semantic labels such as `acoustic', `electronic', or `bright', without timbre magnitude gradation. The SAFE project \cite{stables_safe_2014} proposed a framework for retrieving semantic audio descriptors from plugin metadata, but it does not offer a curated dataset for training generative models. Differentiable digital signal processing approaches such as DDSP \cite{engel_ddsp_2020} focus on interpretable synthesis and parameter conditioning but are not designed to support semantic traversal or descriptor blending in the latent space. Our work advances this line of research by introducing a purpose-built dataset explicitly linking real-world descriptors to timbral magnitudes and validating these associations through both machine learning and perceptual evaluations. This enables structured semantic supervision and high-level timbre control not previously demonstrated in existing audio generation pipelines.

To address this, we introduce the \textbf{Semantic Timbre Dataset} \cite{Cameron_SemanticTimbreDataset_2025}, a curated collection of monophonic electric guitar sounds annotated with 19 distinct timbre descriptors and associated timbre magnitude values. This resource supports the training and evaluation of models for descriptor-based timbre control and semantic audio generation.

%%%%%%%%%%%%%%%%%%%%%%%%
%----------------------%
%%%%%%%%%%%%%%%%%%%%%%%%

\section{Semantic Timbre Dataset}
\label{sec:SemanticTimbreDataset}

The Semantic Timbre Dataset is a systematically constructed collection of 275,310 monophonic electric guitar notes, each labeled with one of 19 semantic timbre descriptors (e.g., `fuzzy', `bright', `shimmering') and a corresponding timbre magnitude value from 0 to 100.

\subsection{Descriptor Selection}
Descriptors were identified via a qualitative content analysis of 72 widely used physical guitar pedals and two virtual plugin suites (Guitar Rig 7 Pro \cite{noauthor_guitar_2023} and Logic Pedalboard \cite{apple_logicpedalboard_2025}). Only the most popularly occurring adjectives and onomatopoeic expressions referring to perceptual timbral qualities were retained. From this analysis, we selected 19 clear, musically relevant descriptors spanning both spectral and temporal timbre effects.

\begin{table}[h]
\centering
\caption{Final 19 timbre descriptors selected for the Semantic Timbre Dataset.}
\begin{tabular}{c|c|c|c}
\toprule
    \multicolumn{2}{c|}{\textbf{Spectral Timbre}} & \multicolumn{2}{c}{\textbf{Temporal Timbre}} \\
\midrule
    \textcolor{red}{DistortionFX} & \textcolor{orange}{FilterFX} & \textcolor{blue}{DynamicsFX} & \textcolor{ForestGreen}{OscillationFX} \\
\midrule
     & \textcolor{orange}{Bright} & \textcolor{blue}{Punchy} & \textcolor{ForestGreen}{Fluttery} \\
    \textcolor{red}{Crunchy} & \textcolor{orange}{Dark} & \textcolor{blue}{Sharp} & \textcolor{ForestGreen}{Jittery} \\
    \textcolor{red}{Crushed} & \textcolor{orange}{Fat} & \textcolor{blue}{Soft} & \textcolor{ForestGreen}{Shimmering} \\
    \textcolor{red}{Dirty} & \textcolor{orange}{Resonant} & \textcolor{blue}{Smooth} & \textcolor{ForestGreen}{Stuttering} \\
    \textcolor{red}{Fuzzy} & \textcolor{orange}{Thin} & \textcolor{blue}{Tight} & \textcolor{ForestGreen}{WahWah} \\
\bottomrule
\end{tabular}
\label{tab:FinalTimbreDescriptors}
\end{table}

Table \ref{tab:FinalTimbreDescriptors} lists the descriptors grouped by their dominant timbral characteristics. Full selection details are available in \cite{Cameron_MPhilThesis_2024}.

\subsection{Audio Synthesis \& Annotation}
We sourced clean monophonic guitar notes from the EGFxSet dataset \cite{egfxset_pedroza_2022} and applied descriptor-specific audio effects using Guitar Rig 7 Pro components. For each descriptor, we varied the effect parameter in steps of 5 (0–100) to generate controlled variations in timbre magnitude. Each resulting audio sample was tagged with the applied timbre descriptor and timbre magnitude, yielding a dense, semantically meaningful dataset suitable for generative modeling.

%%%%%%%%%%%%%%%%%%%%%%%%
%----------------------%
%%%%%%%%%%%%%%%%%%%%%%%%

\section{Semantic Timbre Generation Model}
\label{sec:SemanticTimbreGenerationModel}

To evaluate the dataset’s structure and enable semantic audio generation, we trained an unsupervised variational autoencoder (VAE) on a subset of the Semantic Timbre Dataset. The model learns a latent representation of timbre without requiring conditioning on labels.

\subsection{Model Architecture}
The VAE consists of an encoder–decoder structure implemented in TensorFlow/Keras \cite{chollet_keras_2023}. Input spectrograms of the Semantic Timbre Dataset's sounds are mapped to a 128-dimensional latent space, which was chosen based on perceptual reconstruction quality. The decoder reconstructs log-magnitude spectrograms from these embeddings. Audio is generated via Griffin-Lim phase reconstruction \cite{griffin_signal_1984}. Full implementation details and code are publicly available \cite{Cameron_MPhilThesis_2024,Cameron_SemanticTimbreDataset_Code_2025}.

% Full VAE Model Architecture FIGURE
\begin{figure}[h]
    \centering
    \includegraphics[width=\linewidth]{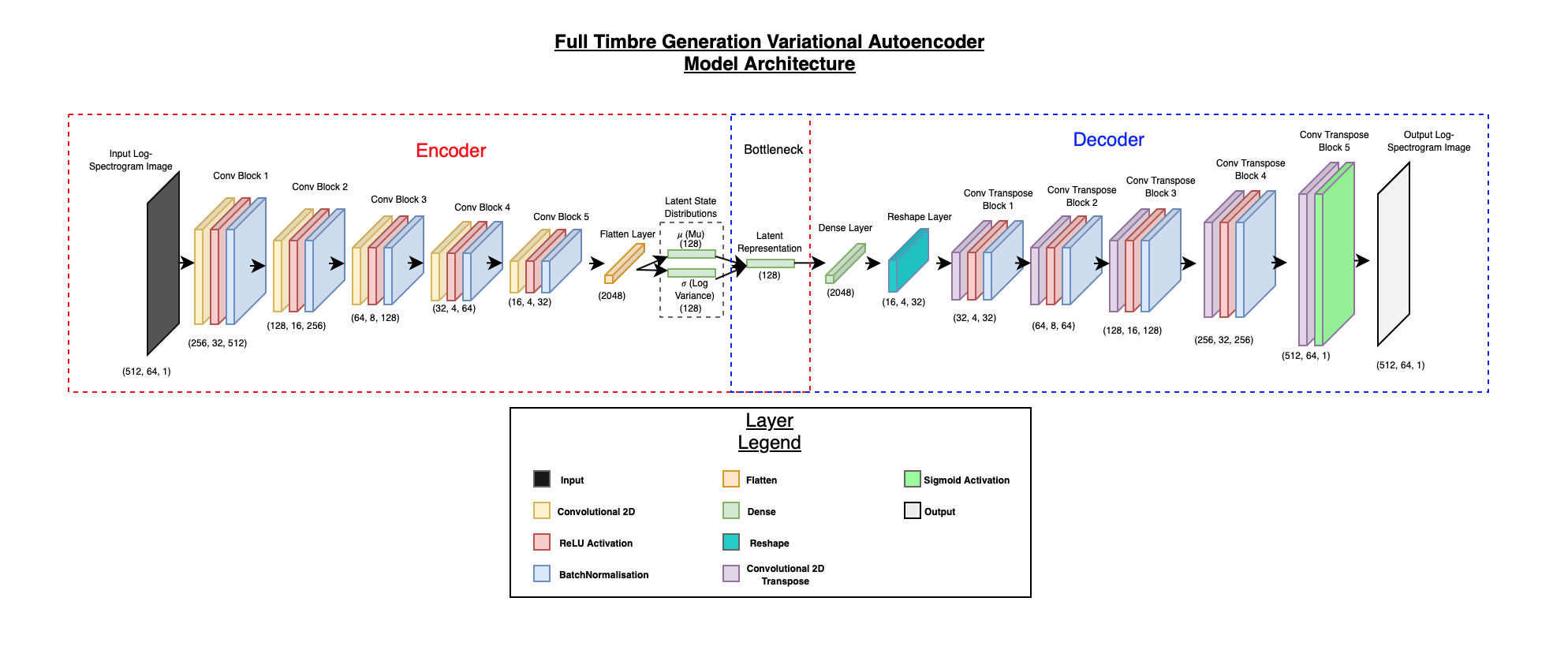}
    \caption{Semantic Timbre Generation VAE model architecture. The encoder maps input spectrograms to a 128D latent space. The decoder reconstructs log-magnitude spectrograms, which are converted back to audio.}
    \label{fig:FullVAEModelArchitecture}
\end{figure}

\subsection{Training Procedure}
The model was trained on 1,771 samples covering E4–D6 notes at 25–100\% descriptor magnitudes, plus clean references \cite{cameron_vaetraindataset_2024}. STFTs were computed with a 1024-point FFT and 512 hop size at 22.05kHz. We used the Adam optimizer \cite{kingma_adam:_2014} (lr = 0.0005), batch size 64, and trained for 300 epochs with a combined reconstruction and KL-divergence loss. Training metrics and spectrogram normalization details are provided in our code repository \cite{Cameron_SemanticTimbreDataset_Code_2025}.

%%%%%%%%%%%%%%%%%%%%%%%%
%----------------------%
%%%%%%%%%%%%%%%%%%%%%%%%

\section{Evaluation \& Results}
\label{sec:EvaluationAndResults}

We evaluated the semantic coherence and generative performance of the VAE through both reconstruction and interpolation tasks using objective metrics and perceptual testing.

\subsection{Timbre Reconstruction}
We generated a test set of E4 notes at four timbre magnitudes (25–100\%) for all 19 descriptors, plus a clean baseline \cite{cameron_vaegenerateddata_2024}. Twenty participants (18 with musical training, 8 guitarists) rated the similarity between original and reconstructed samples using a 5-point Mean Opinion Score (MOS) scale. As seen in Table \ref{tab:TimbreGenerationPercepEvalMOS}, most descriptors achieved MOS $>$ 4.0, with highest scores for `shimmery' (4.96), `fuzz' (4.71), and `thin' (4.86), indicating high perceptual fidelity. Some descriptors involving rapid modulation (`flutter', `stutter', `wahwah') showed lower accuracy, reflecting known limitations of VAEs where they can smooth over fine temporal details \cite{engel_nsynth_2017}.

% MOS TABLE
\begin{table}[h]
\centering
\tiny
\caption{Average Mean Opinion Scores from all 20 participants for the Semantic Timbre Generation System's reconstructions of E4 notes.}
\begin{tabular}{ccccccc}
\toprule
 & & \multicolumn{4}{c}{Timbre Magnitude} & \\
Timbre Group & Timbre Descriptor & 25 & 50 & 75 & 100 & Average MOS \\
\midrule
    \multirow{4}{*}{DistortionFX} & Clean & N/A & N/A & N/A & 4.65 & 4.65 \\
     & Crunch & 4.15 & 3.90 & 4.40 & 4.55 & 4.25 \\
     & Crush & 4.70 & 4.75 & 4.15 & 4.15 & 4.44 \\
     & Dirt & 4.55 & 3.80 & 4.05 & 4.15 & 4.14 \\
     & Fuzz & 4.75 & 4.75 & 4.65 & 4.70 & 4.71 \\
\midrule
    \multirow{4}{*}{FilterFX} & Bright & 4.85 & 4.50 & 4.30 & 4.10 & 4.44 \\
     & Dark & 4.25 & 4.00 & 4.20 & 4.90 & 4.34 \\
     & Fat & 4.10 & 3.45 & 4.35 & 4.30 & 4.05 \\
     & Resonant & 4.75 & 4.95 & 4.80 & 4.80 & 4.83 \\
     & Thin & 4.80 & 4.90 & 4.90 & 4.85 & 4.86 \\
\midrule
    \multirow{4}{*}{DynamicsFX} & Punch & 4.65 & 4.30 & 4.15 & 4.85 & 4.49 \\
     & Sharp & 4.55 & 4.50 & 3.75 & 3.80 & 4.15 \\
     & Smooth & 4.00 & 4.45 & 4.15 & 4.20 & 4.20 \\
     & Soft & 4.55 & 4.70 & 4.85 & 4.85 & 4.74 \\
     & Tight & 3.10 & 2.90 & 2.90 & 2.75 & 2.91 \\
\midrule
    \multirow{4}{*}{OscillationFX} & Flutter & 4.75 & 2.90 & 2.70 & 2.25 & 3.15 \\
     & Jitter & 4.80 & 4.25 & 4.40 & 3.55 & 4.25 \\
     & Shimmer & 5.00 & 5.00 & 5.00 & 4.85 & 4.96 \\
     & Stutter & 3.05 & 2.90 & 2.80 & 2.20 & 2.74 \\
     & WahWah & 4.00 & 2.85 & 2.65 & 2.30 & 2.95 \\
\bottomrule
\end{tabular}
\label{tab:TimbreGenerationPercepEvalMOS}
\end{table}

\subsection{Timbre Interpolation}
To assess semantic blending, we linearly interpolated between latent representations of descriptor pairs (e.g., fuzz–shimmer) \cite{cameron_vaegenerateddata_2024}. Interpolated samples were evaluated using two methods.

\subsubsection{CNN-Based Classification}
\begin{figure}[h]
    \centering
    \includegraphics[width=\linewidth]{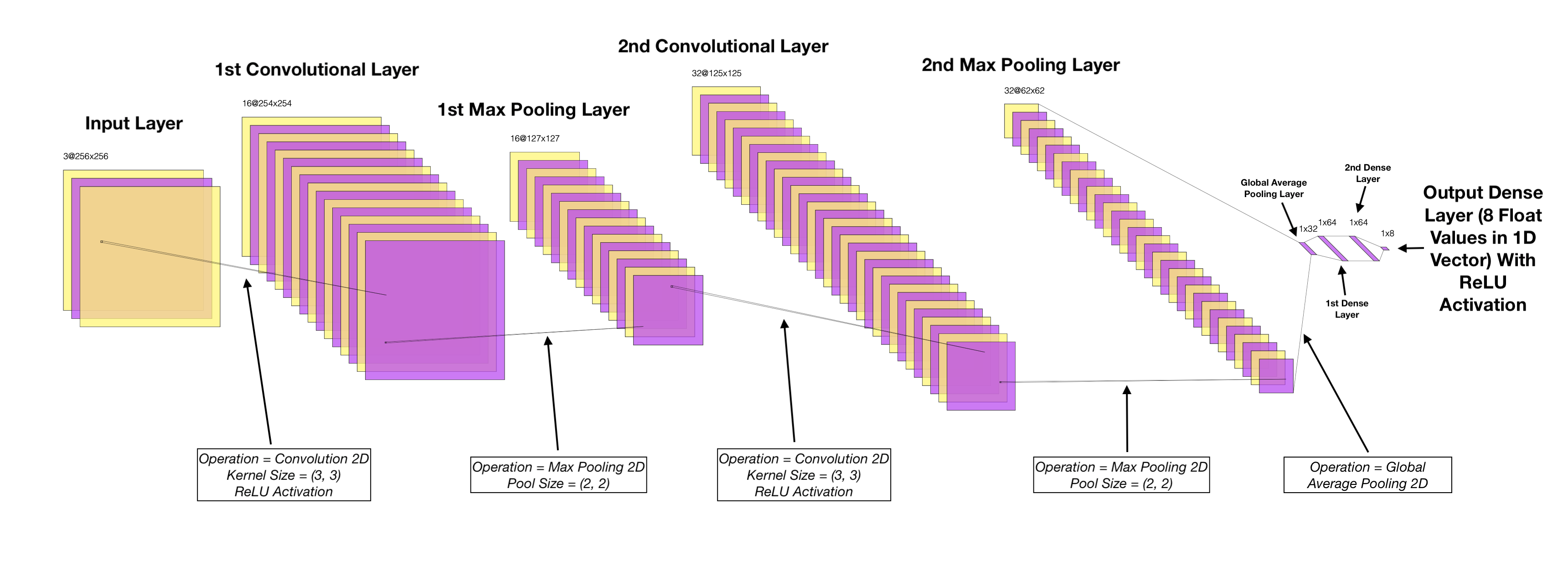}
    \caption{Timbre classifier CNN model architecture.}
    \label{fig:TimbreClassifierModelArchitecture}
\end{figure}

We trained a Convolutional Neural Network (CNN) classifier on spectrograms of descriptor-100\% sounds. The model consists of 3 convolutional blocks followed by two dense layers and a softmax output (see Figure \ref{fig:TimbreClassifierModelArchitecture} for details). It achieved 94.6\% validation accuracy and was used to evaluate semantic accuracy of interpolated sounds.

\begin{figure}[h]
    \centering
    \begin{subfigure}[t]{.2\textwidth}
    \centering
    \includegraphics[width=\linewidth]{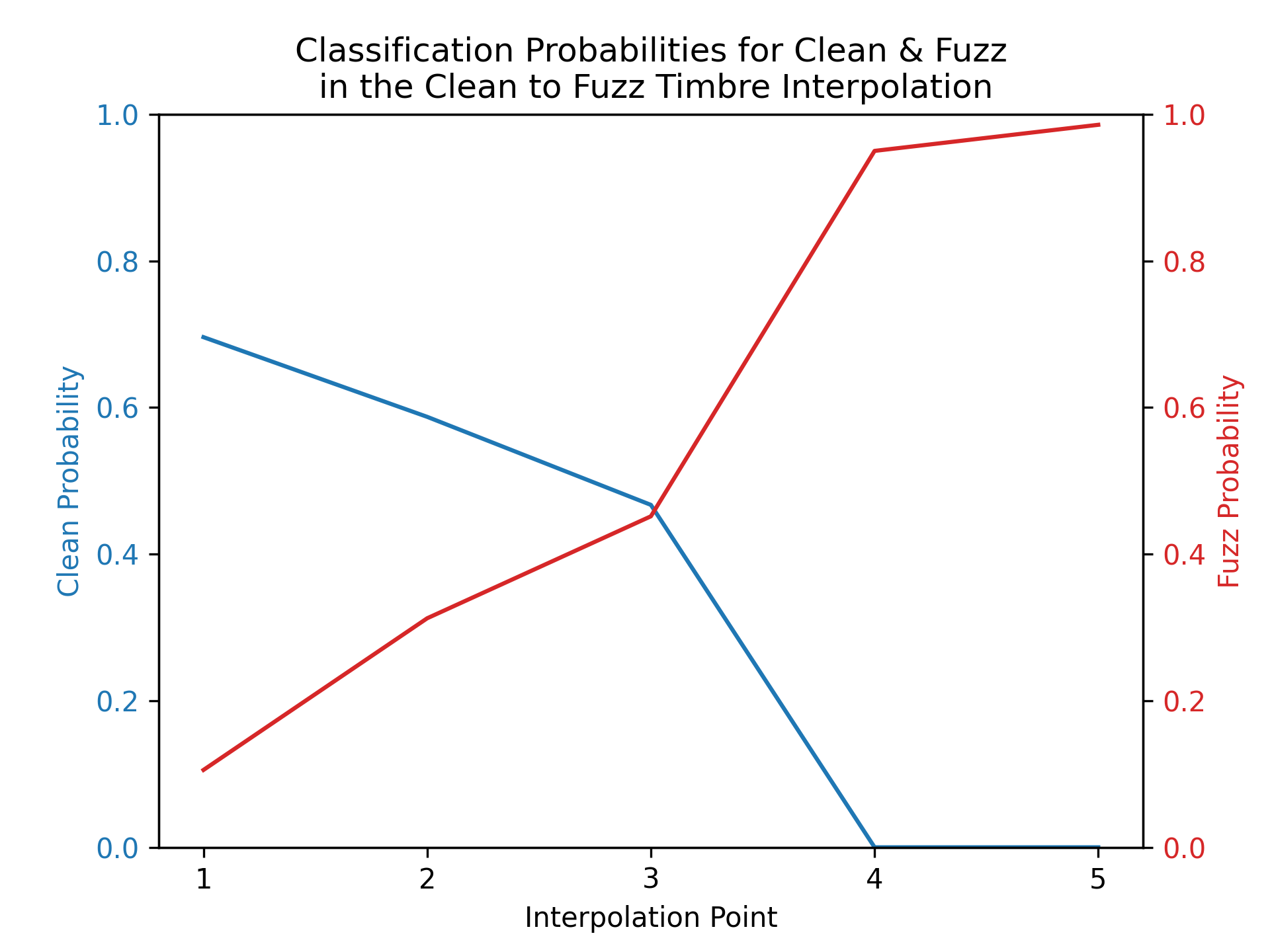}
    \caption{Clean-Fuzz}
    \label{fig:ClassCleanToFuzz}
    \end{subfigure}
    \begin{subfigure}[t]{.2\textwidth}
    \centering
    \includegraphics[width=\linewidth]{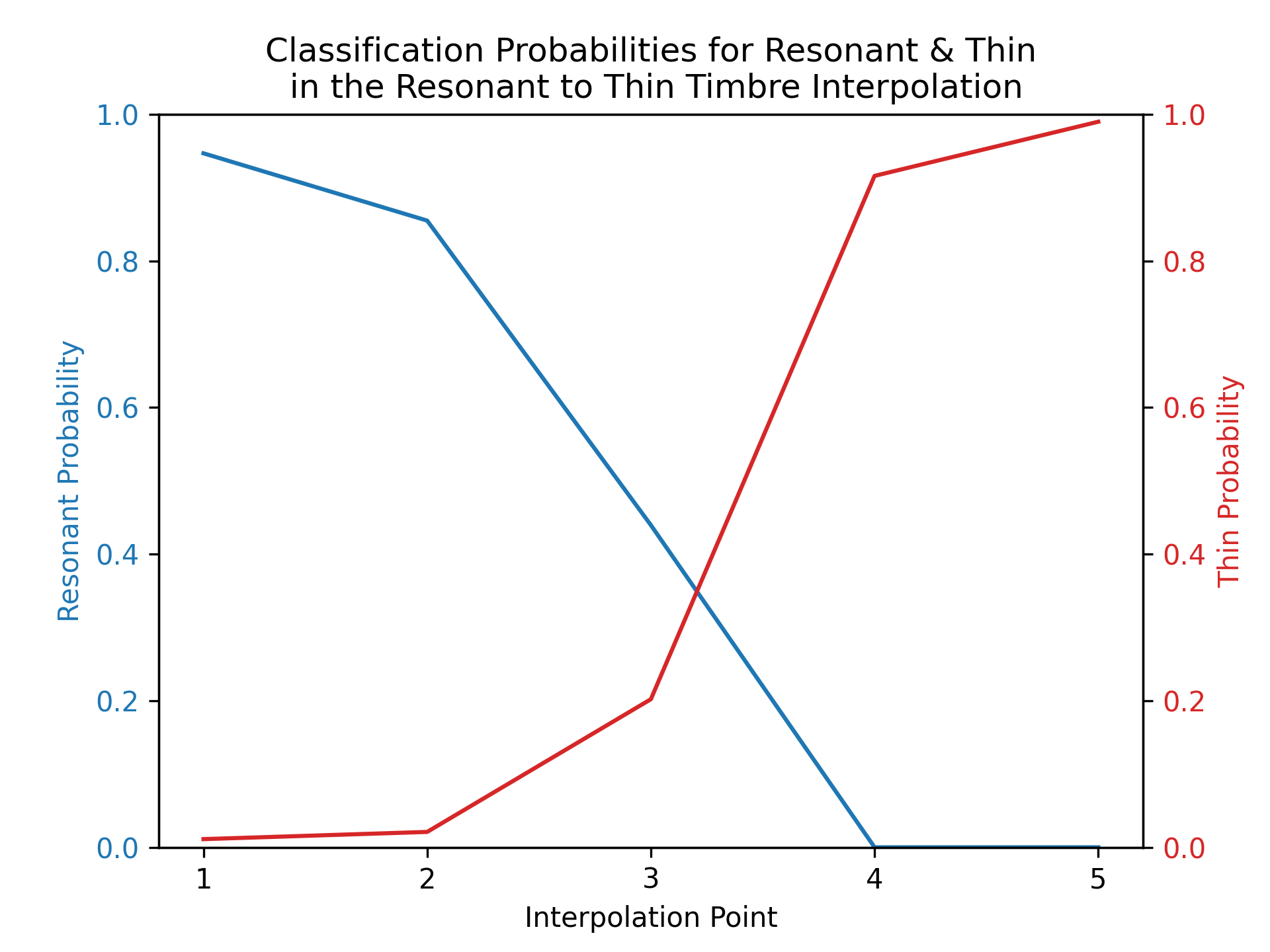}
    \caption{Resonant-Thin}
    \label{fig:ClassResonantToThin}
    \end{subfigure}\hspace{2em}
    \begin{subfigure}[t]{.2\textwidth}
    \centering
    \includegraphics[width=\linewidth]{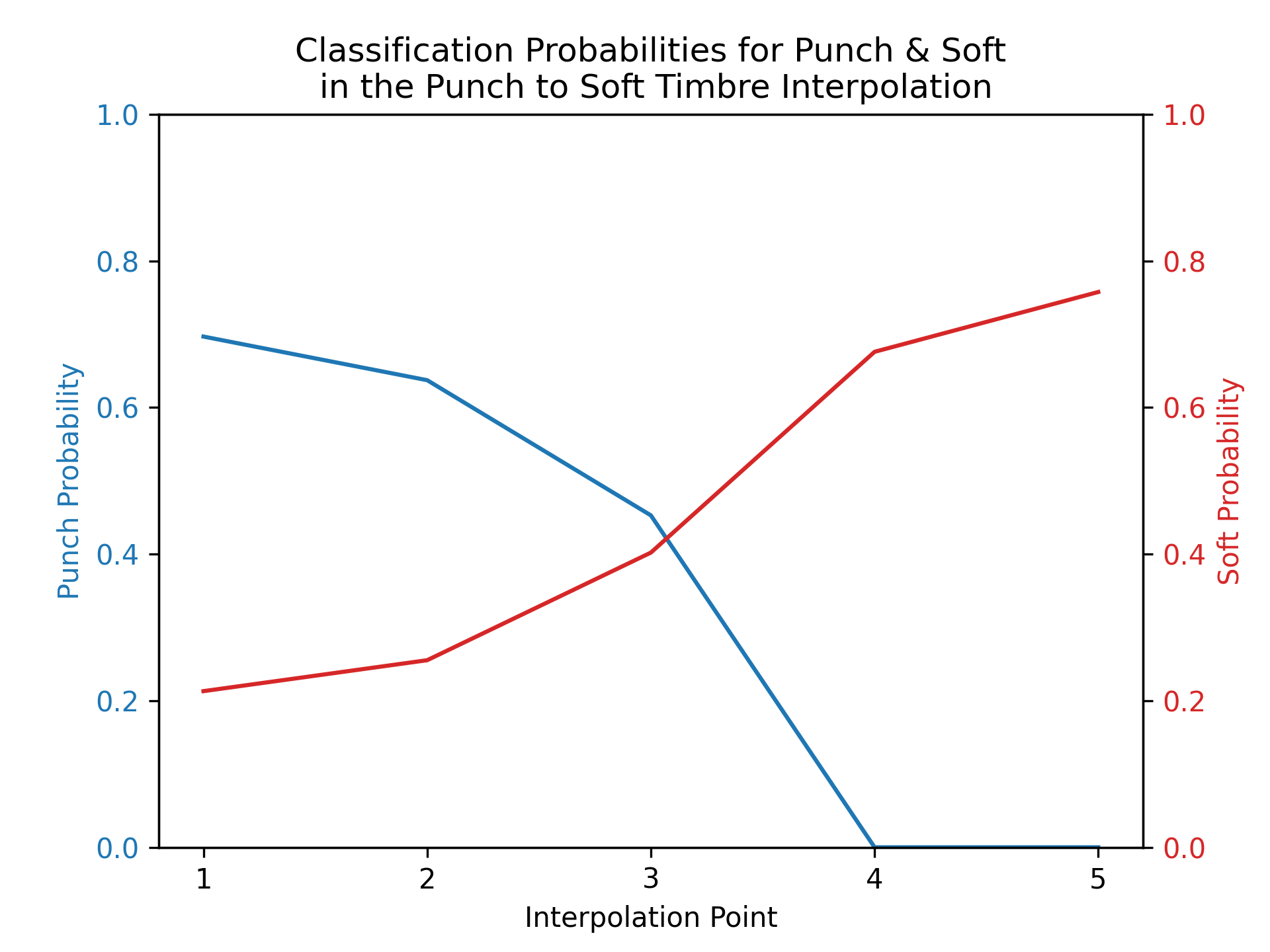}
    \caption{Punch-Soft}
    \label{fig:ClassPunchToSoft}
    \end{subfigure}
    \begin{subfigure}[t]{.2\textwidth}
    \centering
    \includegraphics[width=\linewidth]{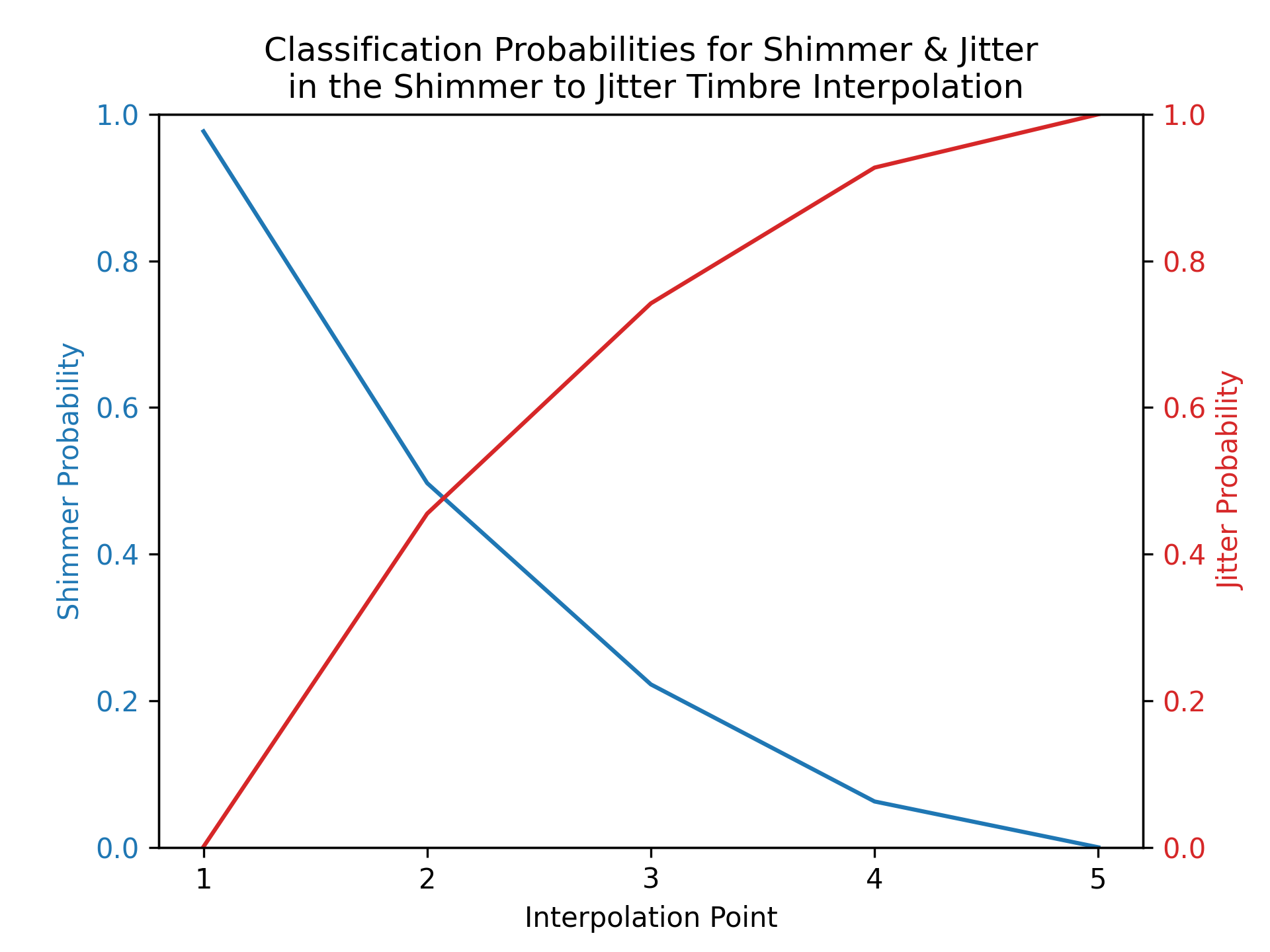}
    \caption{Shimmer-Jitter}
    \label{fig:ClassShimmerToJitter}
    \end{subfigure}
    \caption{Classification probabilities for interpolations between the top 2 descriptors for each timbre group.}
    \label{fig:TimbreInterpClassPlotsEachGroup}
\end{figure}

Figure \ref{fig:TimbreInterpClassPlotsEachGroup} shows the classifier’s probabilities for 5-point interpolations between the timbre descriptor pairs obtained via the top-2 performing timbre descriptor reconstructions from each timbre group in the previously mentioned timbre reconstruction evaluation. Table \ref{tab:TimbreGenerationPercepEvalMOS} shows that the top-2 timbre descriptors for each timbre group are: DistortionFX - `Clean' \& `Fuzz', FilterFX - `Resonant' \& `Thin', DynamicsFX - `Punch' \& `Soft', OscillationFX - `Jitter' \& `Shimmer'. This results in 4 5-point interpolations (Clean-Fuzz, Resonant-Thin, Punch-Soft, and Shimmer-Jitter).

% TOP TWO DESCRIPTOR INTERPOLATION PLOTS

\begin{figure}[h]
    \centering
    \begin{subfigure}[t]{.2\textwidth}
    \centering
    \includegraphics[width=\linewidth]{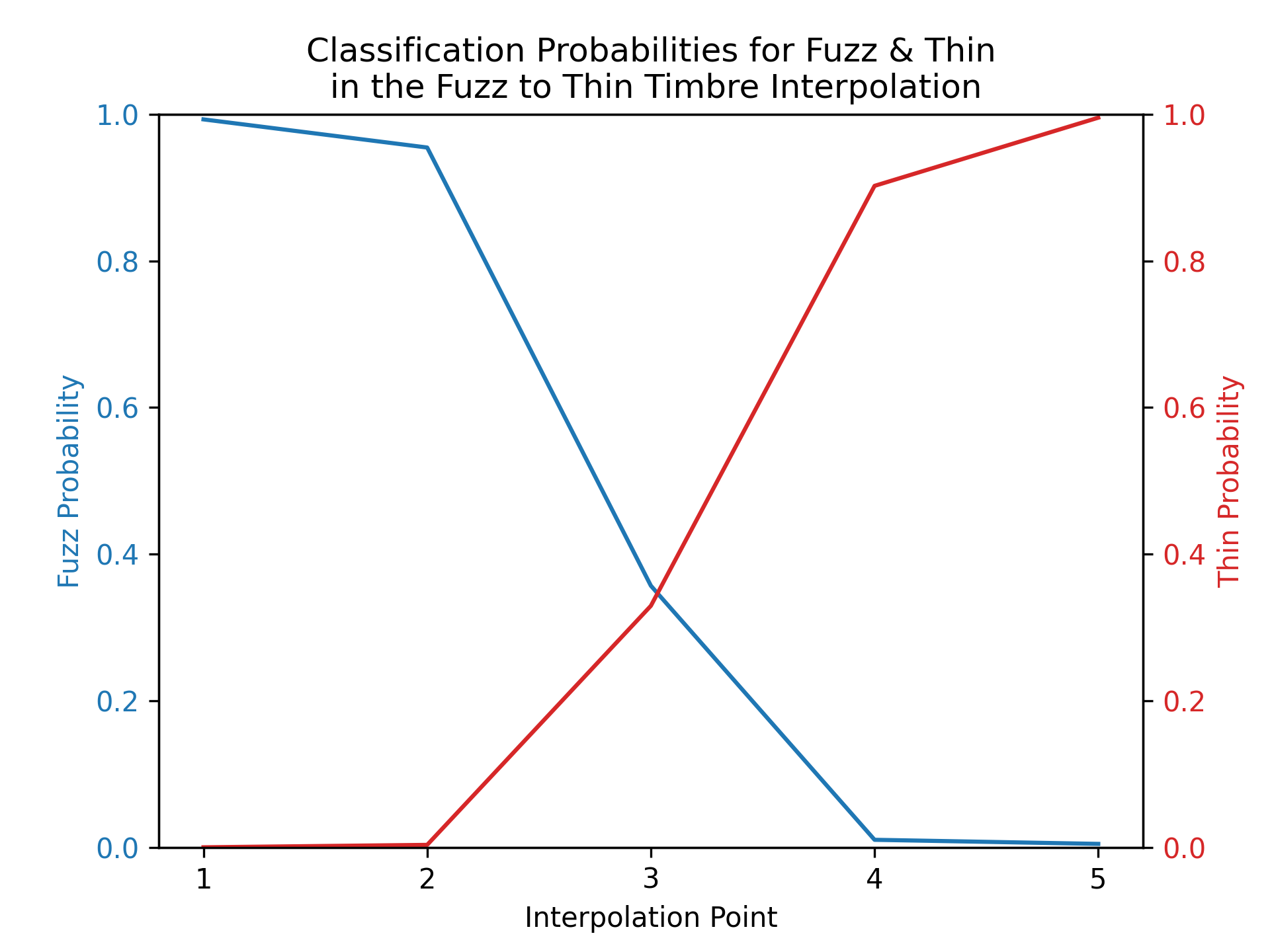}
    \caption{Fuzz-Thin}
    \label{fig:ClassFuzzToThin}
    \end{subfigure}
    \begin{subfigure}[t]{.2\textwidth}
    \centering
    \includegraphics[width=\linewidth]{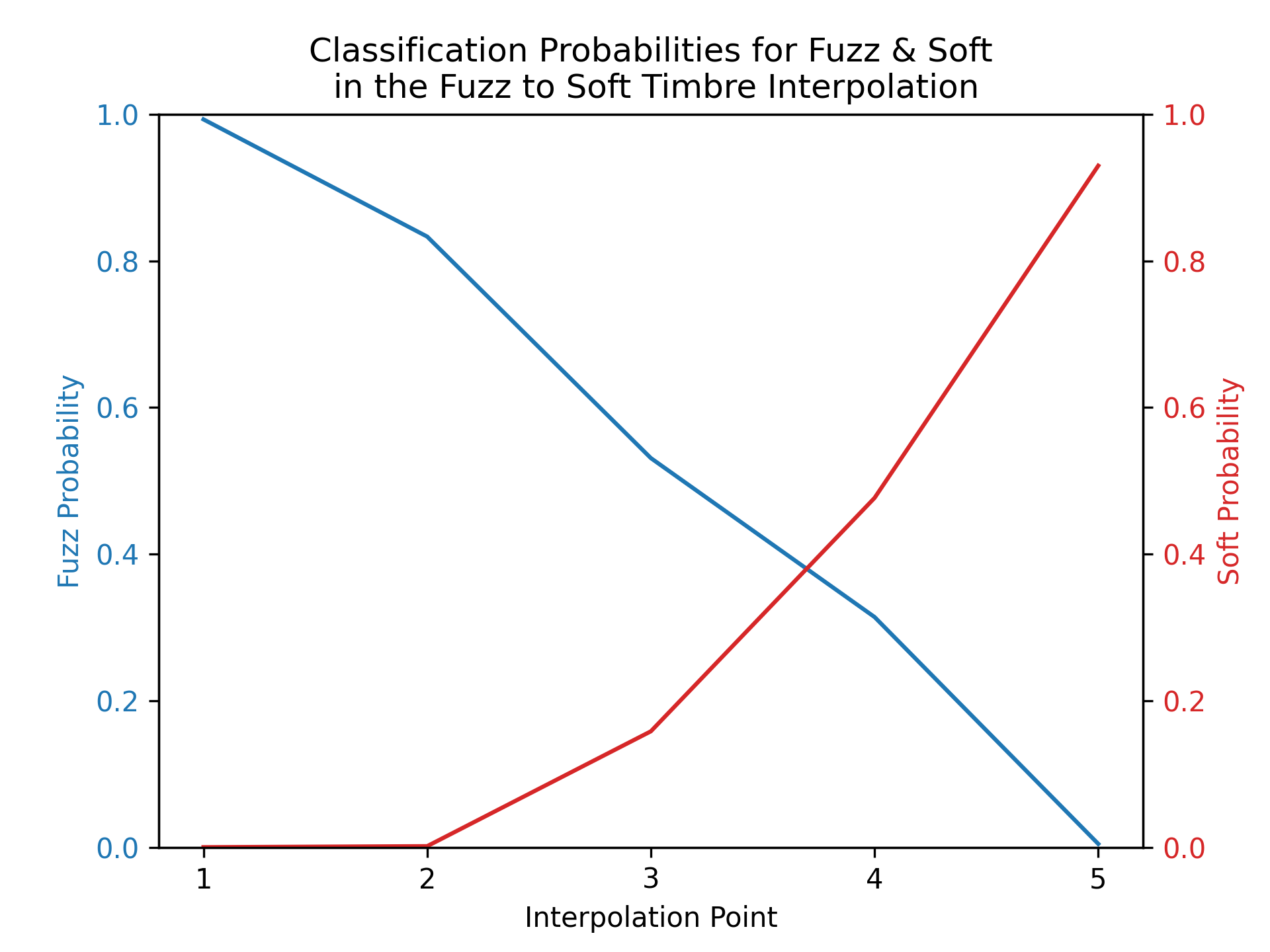}
    \caption{Fuzz-Soft}
    \label{fig:ClassFuzzToSoft}
    \end{subfigure}\hspace{2em}
    \begin{subfigure}[t]{.2\textwidth}
    \centering
    \includegraphics[width=\linewidth]{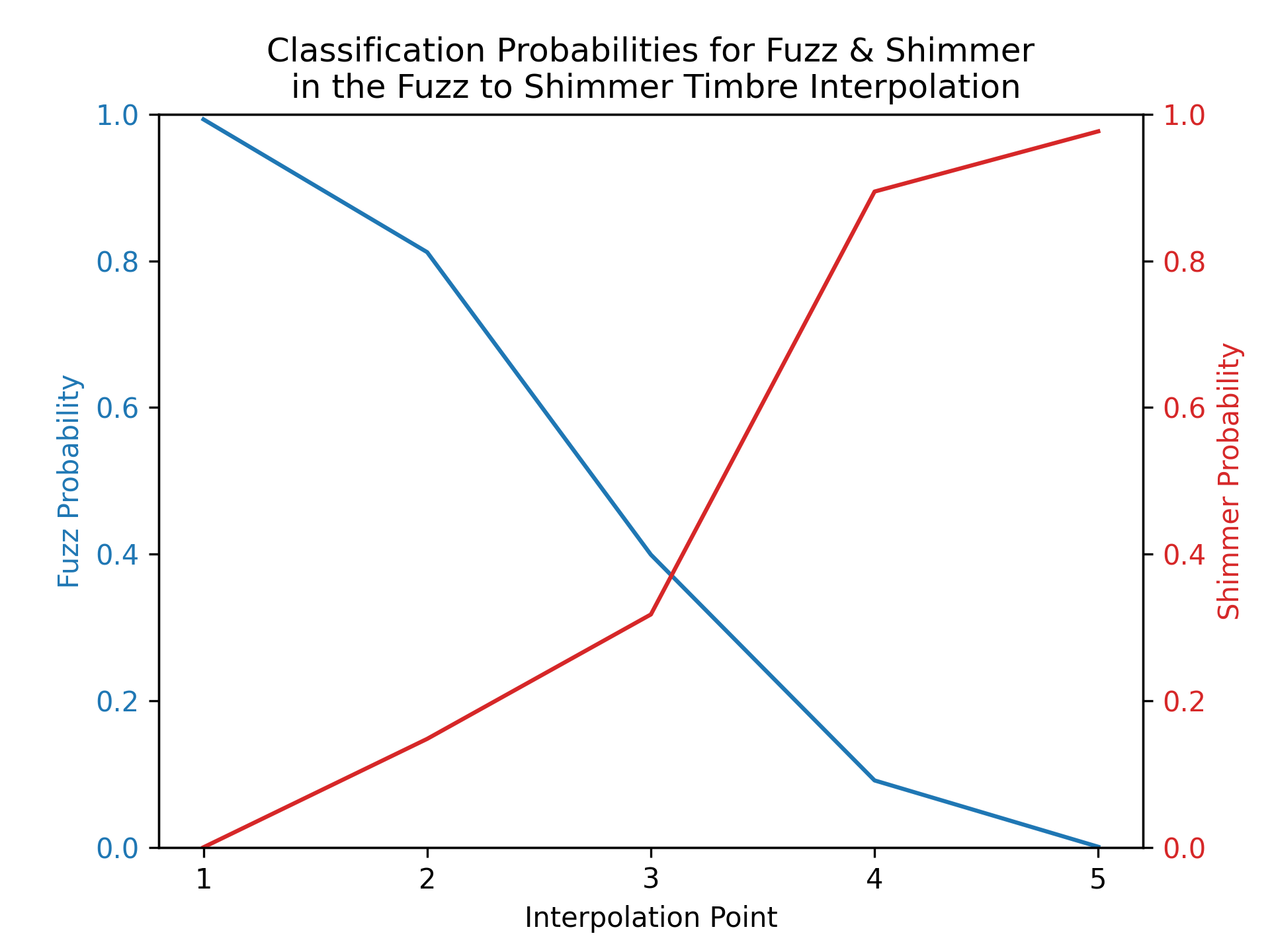}
    \caption{Fuzz-Shimmer}
    \label{fig:ClassFuzzToShimmer}
    \end{subfigure}\hspace{2em}
    \begin{subfigure}[t]{.2\textwidth}
    \centering
    \includegraphics[width=\linewidth]{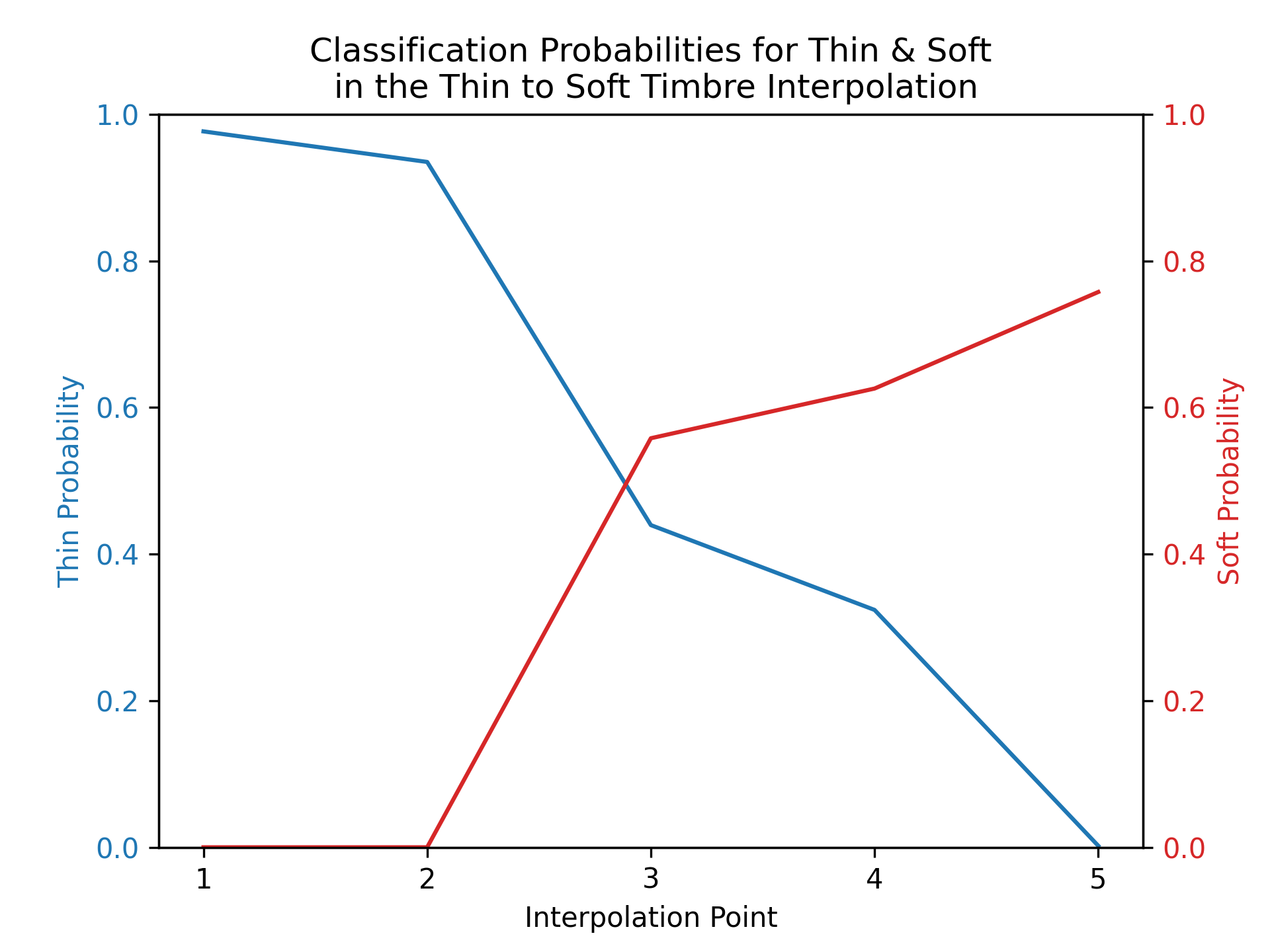}
    \caption{Thin-Soft}
    \label{fig:ClassThinToSoft}
    \end{subfigure}\hspace{2em}
    \begin{subfigure}[t]{.2\textwidth}
    \centering
    \includegraphics[width=\linewidth]{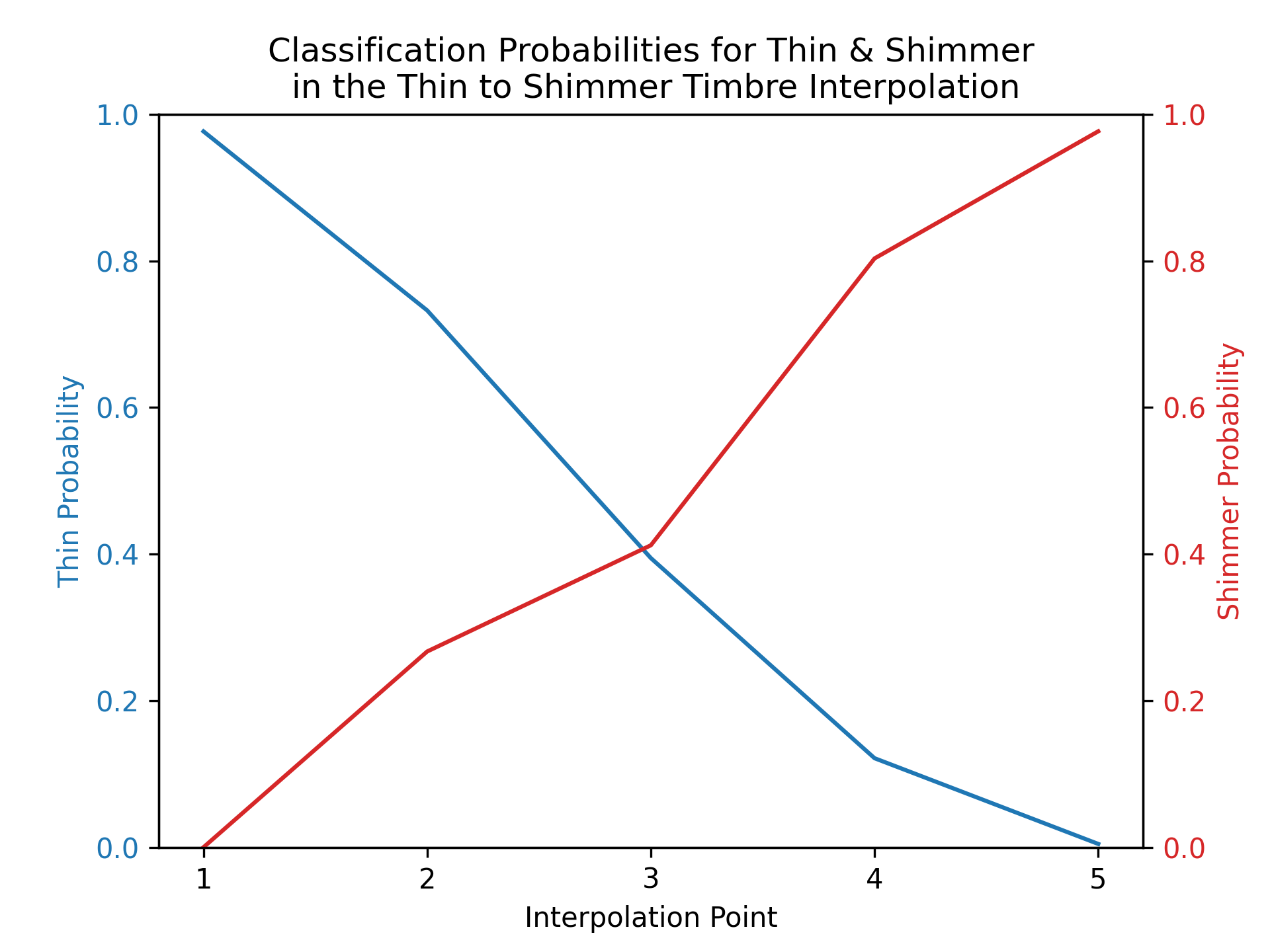}
    \caption{Thin-Shimmer}
    \label{fig:ClassThinToShimmer}
    \end{subfigure}
    \begin{subfigure}[t]{.2\textwidth}
    \centering
    \includegraphics[width=\linewidth]{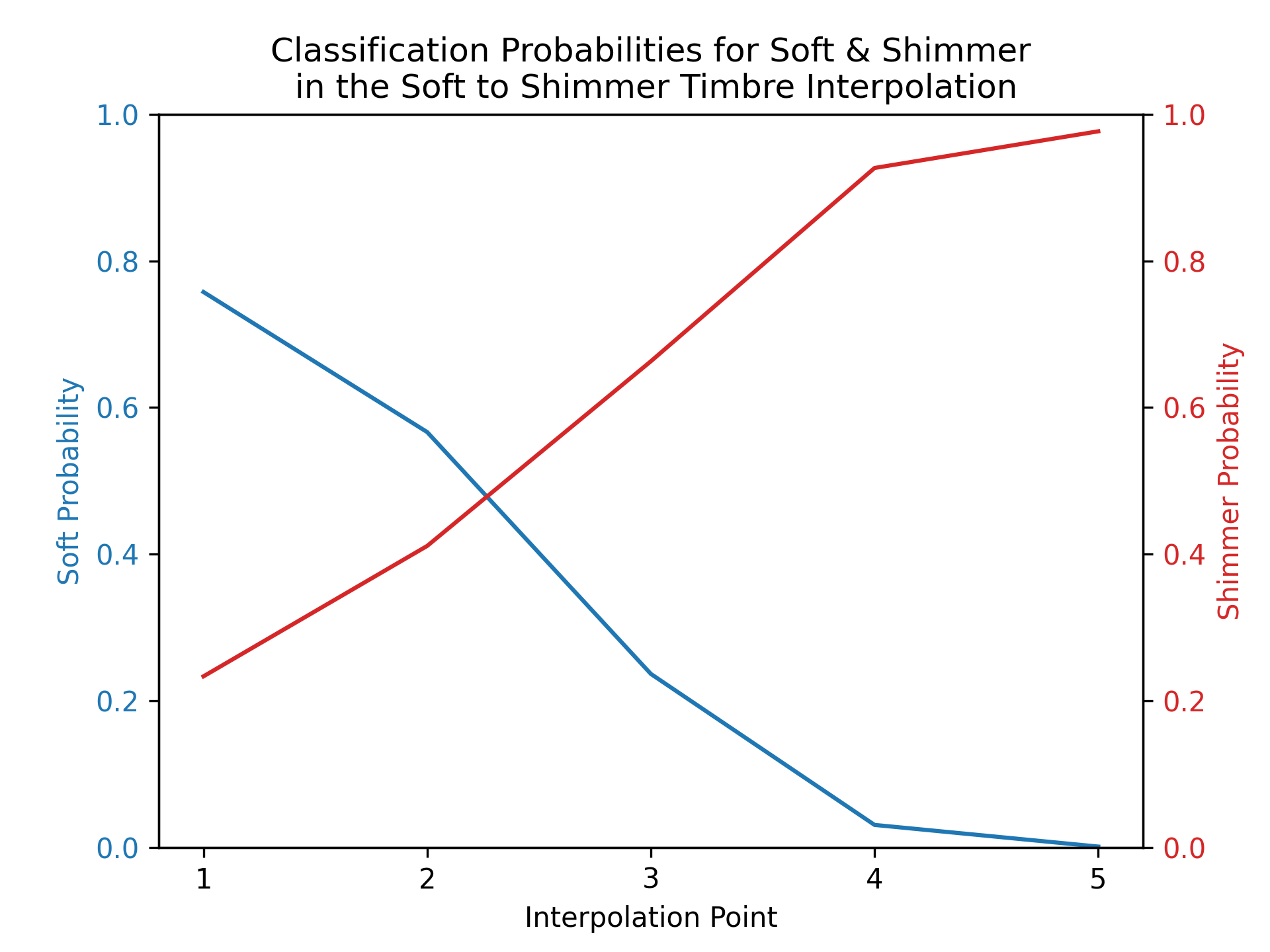}
    \caption{Soft-Shimmer}
    \label{fig:ClassSoftToShimmer}
    \end{subfigure}
    \caption{Classification probabilities for interpolations between the top descriptors of each timbre group.}
    \label{fig:TimbreInterpClassPlotsTop}
\end{figure}

Figure \ref{fig:TimbreInterpClassPlotsTop} shows the classifier’s probabilities for 5-point interpolations between the timbre descriptor pairs derived via all pair combinations of the top performing timbre descriptor reconstructions from all timbre groups in the previously mentioned timbre reconstruction evaluation. Table \ref{tab:TimbreGenerationPercepEvalMOS} shows that the top timbre descriptors for each timbre group are: DistortionFX - `Fuzz', FilterFX - `Thin', DynamicsFX - `Soft', OscillationFX - `Shimmer'. This results in 6 5-point interpolations (Fuzz-Thin, Fuzz-Soft, Fuzz-Shimmer, Thin-Soft, Thin-Shimmer, and Soft-Shimmer).

% TOP DESCRIPTOR FOR EACH GROUP INTERPOLATION PLOTS

These CNN-based classification results objectively demonstrated semantic interpolation. The classifier identified the start and target descriptors at interpolation endpoints, with probabilities decreasing toward midpoint interpolations, thus validating the semantic transitions encoded in the VAE’s latent space.

\subsubsection{Perceptual Interpolation Evaluation}
\begin{figure}[h]
    \centering
    \includegraphics[width=0.8\linewidth]{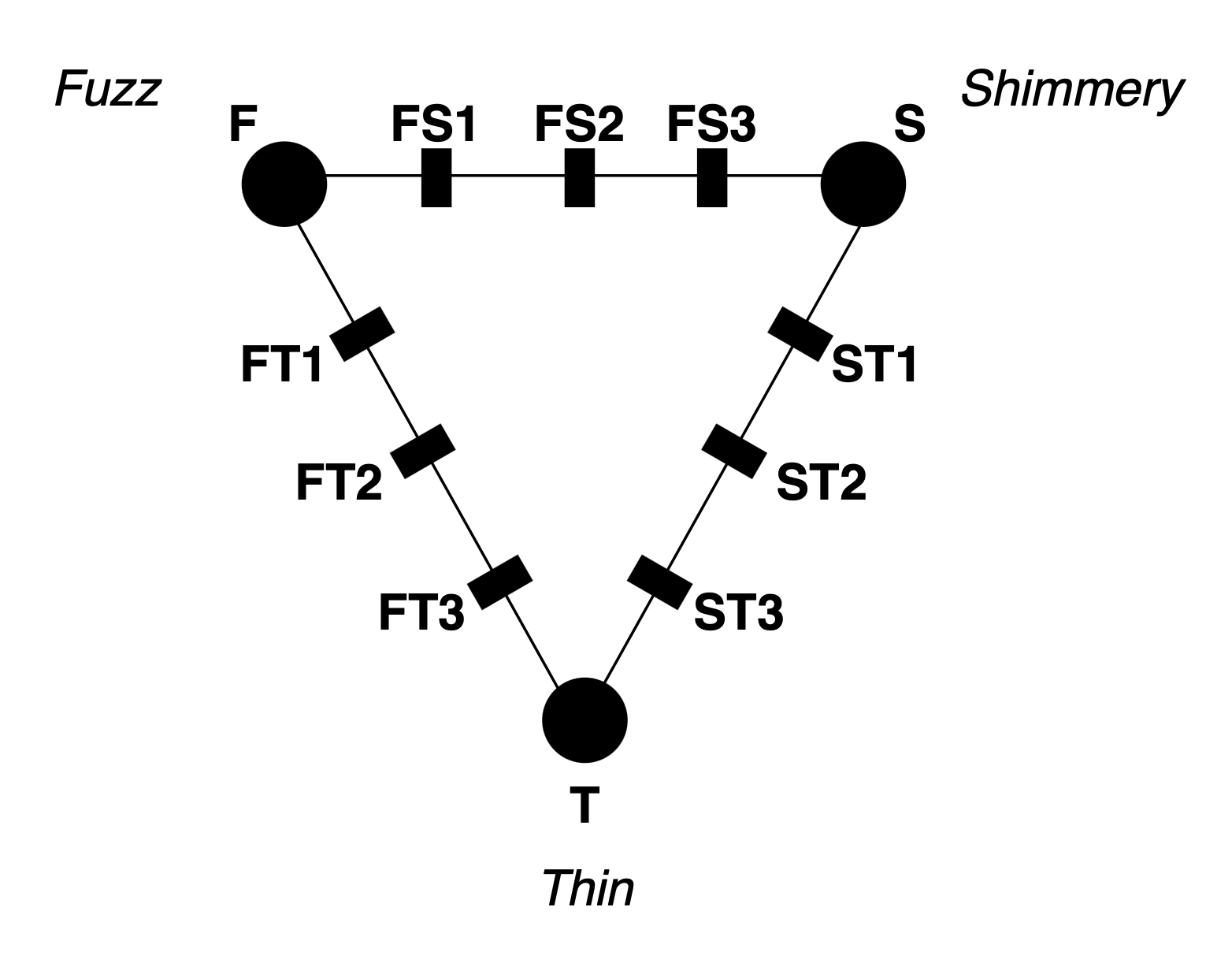}
    \caption{The triangular template used by participants to order interpolated sounds.}
    \label{fig:TriangleTemplate}
\end{figure}

For a robust perceptual assessment of interpolation quality, the same 20 participants from the reconstruction evaluation were tasked with perceptually organizing interpolated audio samples relative to the original descriptor endpoints. Participants positioned interpolated sounds resulting from 5-point interpolations between Fuzz-Shimmer, Shimmer-Thin, and Fuzz-Thin within a triangular space defined by the Semantic Timbre Dataset's original `Fuzz’, `Shimmer’, and `Thin’ sounds, reflecting perceived timbral distances. This setup, which is visualised in Figure \ref{fig:TriangleTemplate}, evaluated the perceptual coherence and semantic accuracy of the interpolated sounds.

% INTERPOLATION DIAGRAM FIGURE

% CONFUSION MATRIX FIGURE
\begin{figure}[h]
    \centering
    \includegraphics[width=0.8\linewidth]{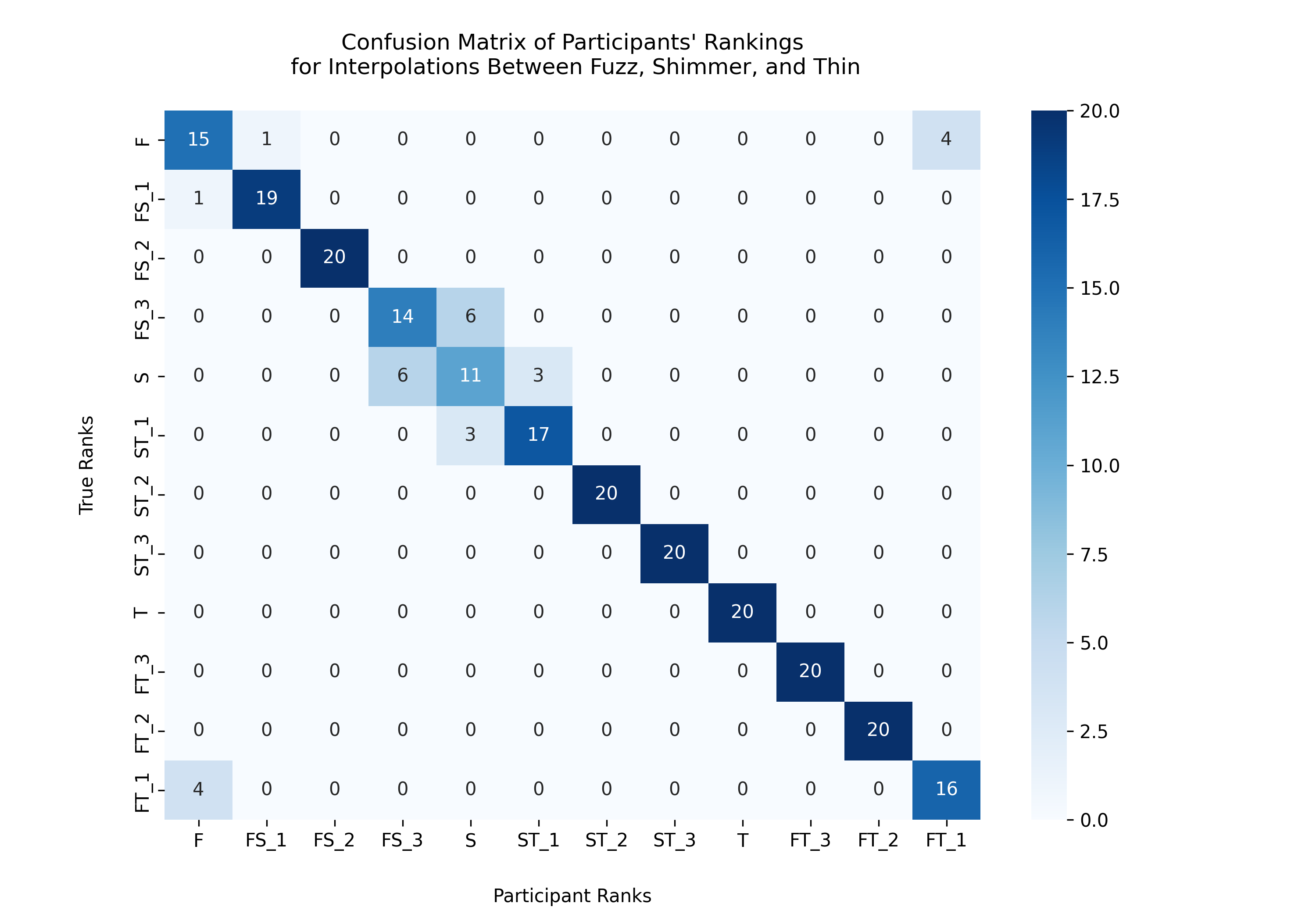}
    \caption{Participants’ placements for interpolated audio samples. Refer to Figure \ref{fig:TriangleTemplate} for class locations on the experiment’s triangle.}
    \label{fig:ConfusionMatrix}
\end{figure}

Participants effectively organized interpolated sounds according to perceptual timbral distances, correctly placing most sounds and demonstrating clear semantic coherence. Figure \ref{fig:ConfusionMatrix} displays where participants placed sounds on the triangular template shown in Figure \ref{fig:TriangleTemplate} versus the sounds’ true placements via a confusion matrix. Misplacements were minimal and typically limited to adjacent interpolation points, reflecting minor perceptual ambiguities particularly around ‘fuzz’ and ‘shimmer’. To statistically assess participants’ perceptual accuracy, we calculated a Kendall’s Tau correlation coefficient $\tau = 0.879$ between the true positions of interpolated sounds and the aggregated positions averaged across the 20 participants.

\begin{figure}[h]
    \centering
    \includegraphics[width=0.8\linewidth]{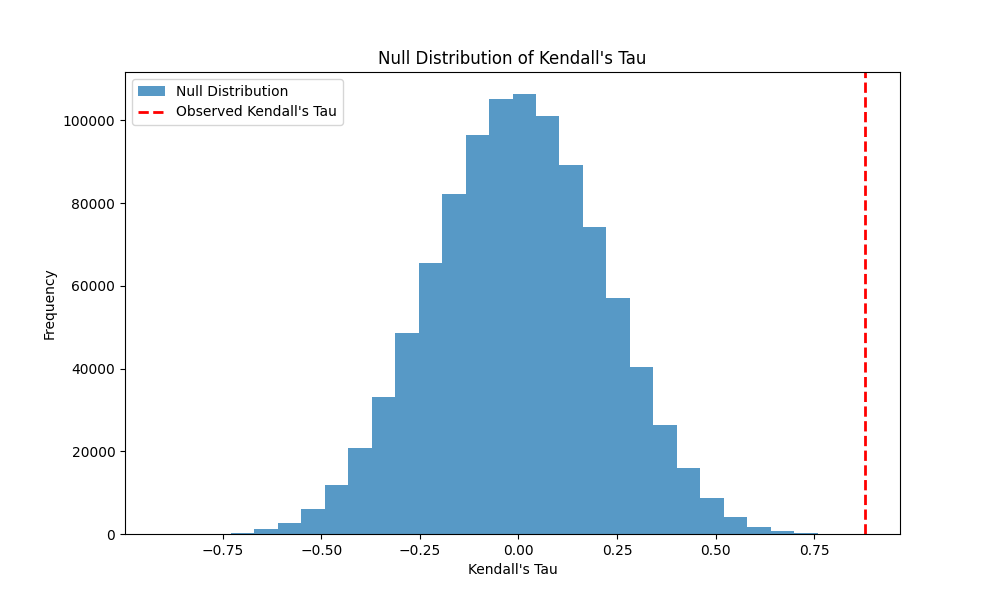}
    \caption{Null distribution of Kendall’s Tau correlation values from 1,000,000 permutations of true rankings. The observed $\tau$ = 0.879 from participant rankings is highlighted, demonstrating strong statistical support ($p < 0.001$) for perceptual coherence of VAE-generated timbre interpolations.}
    \label{fig:KendallsTauDistribution}
\end{figure}

We then tested the null hypothesis that participant rankings of the interpolated sounds were no better than random through a permutation test, randomly permuting the true ranks 1,000,000 times to build a null distribution of Kendall’s Tau values. Comparing the observed $\tau = 0.879$ against this distribution resulted in a significant p-value $<$ 0.001, indicating that participant placements were reliably better than chance. Figure \ref{fig:KendallsTauDistribution} illustrates this distribution alongside the observed Kendall’s Tau value, confirming strong statistical support for perceptual coherence in semantic timbre interpolation.

% NULL DISTRIBUTION FIGURE

These findings validate the dataset’s structure and demonstrate the VAE’s capacity to support descriptor-based control and interpolation.

%%%%%%%%%%%%%%%%%%%%%%%%
%----------------------%
%%%%%%%%%%%%%%%%%%%%%%%%

\section{Conclusion}
\label{sec:Conclusion}

This paper introduces the Semantic Timbre Dataset, a novel and rigorously annotated audio resource explicitly created for advancing semantic timbre control and generative modeling. Our dataset bridges critical gaps in existing audio data resources by systematically associating detailed semantic descriptors, derived from comprehensive qualitative analysis of guitar effects units, with monophonic electric guitar sounds. Evaluations using an unsupervised VAE demonstrated robust semantic timbre reconstruction and interpolation capabilities, validating the intrinsic semantic coherence of the dataset. These results establish a promising benchmark for future semantic audio modeling research.

%%%%%%%%%%%%%%%%%%%%%%%%
%----------------------%
%%%%%%%%%%%%%%%%%%%%%%%%

\clearpage
\vfill\pagebreak

\bibliographystyle{IEEEbib}
\bibliography{refs}

\section{Compliance with Ethical Standards}
User studies involving human participants were performed in line with the principles of the University of Cambridge. Approval was granted by the Ethics Committee of the School of Technology, University of Cambridge during April 2024.

\end{document}